\documentclass{pasa}%

\usepackage{graphicx}
\usepackage{subfig}
\usepackage{natbib}

\newcommand{\msun}{\ifmmode\mbox{M}_{\odot}\else$\mbox{M}_{\odot}$\fi}
\newcommand{\arcdeg}{\mbox{$^\circ$}}
\newcommand{\degr}{\arcdeg}
\newcommand{\fs}{\mbox{$.\!\!^{\mathrm s}$}}
\newcommand\fdg{\mbox{$.\!\!^\circ$}}
\newcommand{\farcs}{\mbox{$.\!\!^{\prime\prime}$}}
\newcommand{\maspy}{$\rm mas~yr^{-1}$}

\newcommand{\gx}{GX~17$+$2}
\newcommand{\cyg}{Cyg~X$-$2}
\newcommand{\cen}{Cen~X$-$4}
\newcommand{\uua}{4U~0919$-$54}

\newcommand{\xb}{XB~2129$+$47}

\newcommand{\sax}{SAX~J1808.4$-$3658}

\newcommand{\multilinecomment}[1]{}

\title[Gaia astrometry of type I X-ray bursters]{Gaia EDR3 parallaxes of type I X-ray bursters and their implications on the models of type I X-ray bursts: a generic approach to the Gaia parallax zero-point and its uncertainty}

\author[Ding, Deller \& Miller-Jones]{Hao Ding$^{1,2}$\thanks{haoding@swin.edu.au}, Adam T. Deller$^{1,2}$ and James C. A. Miller-Jones$^3$
\affil{$^1$Centre for Astrophysics and Supercomputing, Swinburne University of Technology, VIC 3122, Australia}%
\affil{$^2$ARC Centre of Excellence for Gravitational Wave Discovery (OzGrav)}
\affil{$^3$International Centre for Radio Astronomy Research---Curtin University, WA 6845, Australia}
}

\jid{PASA}
\doi{10.1017/pas.\the\year.xxx}
\jyear{\the\year}

\usepackage{aas_macros}
\usepackage[]{hyperref} 
\hypersetup{colorlinks,citecolor=blue,linkcolor=blue,urlcolor=blue}

\begin{document}

\begin{frontmatter}
\maketitle

\begin{abstract}
Light curves of photospheric radius expansion (PRE) bursts, a subset of type I X-ray bursts, have been used as standard candles to estimate the ``nominal PRE distances'' for 63\% of PRE bursters (bursters), assuming PRE burst emission is spherically symmetric.
Model-independent geometric parallaxes of bursters provide a valuable chance to test models of PRE bursts (PRE models), and can be provided in some cases by Gaia astrometry of the donor stars in bursters.
We searched for counterparts to 115 known bursters in the Gaia Early Data Release 3, and confirmed 4 bursters with Gaia counterparts that have detected ($>3\,\sigma$, prior to zero-point correction) parallaxes. 
We describe a generic approach to the Gaia parallax zero point as well as its uncertainty 
using an ensemble of Gaia quasars individually determined for each target. 
Assuming the spherically symmetric PRE model is correct, we refined the resultant nominal PRE distances of three bursters (i.e. \cen, \cyg\ and \uua), and put constraints on their compositions of the nuclear fuel powering the bursts. Finally, we describe a method for testing the correctness of the spherically symmetric PRE model using parallax measurements, and provide preliminary results.
\end{abstract}

\end{frontmatter}

\section{INTRODUCTION}
\label{sec:intro}

A type I X-ray burst is rapid nuclear burning on the surface of a neutron star (NS) \citep[e.g.][]{Lewin93}. 
Type I X-ray bursts happen when the accumulated ``nuclear fuel'' accreted from the donor star onto the surface of the NS is sufficiently hot. 
In some cases, the radiation pressure becomes large enough to overcome the gravity of the NS, causing the photosphere to expand. This is followed by fallback of the photosphere after extending to its outermost point. This process of expansion and contraction of the photosphere is well characterised by double-peaked soft X-ray light curves. These particular type I X-ray bursts are called photospheric radius expansion (PRE) bursts.

According to \citet{Galloway20}, there are hitherto 115 recognized PRE bursters (see \url{https://burst.sci.monash.edu/sources}), which are all low mass X-ray binaries (LMXBs).
PRE bursts can serve as standard candles \citep{Basinska84} based on the assumption that the luminosity of PRE bursts stays at the Eddington limit throughout the photospheric expansion and contraction. We refer to the distances estimated in this way as PRE distances. 
Using the standard assumption of spherically symmetric PRE burst emission, \citet{Galloway20} constrained PRE distances for 73 PRE bursters (23 of which only have upper limits of PRE distances, see Table~8 of \citealp{Galloway20}) based on the theoretical Eddington luminosity calculated by \citet{Lewin93}. 
Hereafter, we refer to this spherically symmetric model as the simplistic PRE model, and the simplistic-PRE-model-based PRE distances as nominal PRE distances.
Besides PRE models, a Bayesian framework has been recently developed to infer parameters, including the distance and the composition of nuclear fuel on the NS surface, for the type I X-ray burster \sax\ \citep{Goodwin19}, by matching the burst observables (such as burst flux and recurrence time) of non-PRE type I X-ray bursts with the prediction from a burst ignition model \citep{Cumming00}. 

At least 43\% of PRE bursters registered nominal PRE distances as their best constrained distances.
The great usefulness of the simplistic PRE model calls for careful examination of its validity. There are several uncertainties in the simplistic PRE model. Firstly, according to \citet{Lewin93}, the Eddington luminosity measured by an observer at infinity
\begin{equation}
\label{eq:Eddington_luminosity}
L_{\mathrm{Edd},\infty} \propto 
\frac{M_\mathrm{NS}}{1+z(R_\mathrm{P})} \frac{1}{1+X}
\end{equation}
depends on the NS mass $M_\mathrm{NS}$ and the gravitational redshift $z(R_\mathrm{P})=(1-2G_\mathrm{N} M_\mathrm{NS}/c^2 R_\mathrm{P})^{-1/2}-1$ (where $G_\mathrm{N}$ and $c$ stand for Newton's gravitational constant and speed of light in vacuum, respectively) at the photosphere radius $R_\mathrm{P}$. To a greater degree, $L_{\mathrm{Edd},\infty}$ hinges on the hydrogen mass fraction $X$ of the nuclear fuel on the NS surface at the time of the PRE burst: hydrogen-free nuclear fuel corresponds to a $L_{\mathrm{Edd},\infty}$ 1.7 times higher than nuclear fuel of cosmic abundances \citep{Lewin93}. Secondly, the method assumes spherically symmetric emission. 
On one hand, PRE bursts per se are not necessarily spherically symmetric given that the NS in an LMXB is spinning and accretes via an accretion disk.
On the other hand, even if PRE bursts are initially isotropic (assuming the nuclear fuel has spread evenly on the NS surface), propagation effects (such as the reflection from the surrounding accretion disk) would still potentially lead to anisotropy of PRE burst emission.
Thirdly, for each PRE burster, the peak fluxes of its recognized PRE bursts vary, typically by 13\% \citep{Galloway08}. 

Provided the uncertainties of the simplistic PRE model as well as the resultant nominal PRE distances, independent distance measurements for PRE bursters hold the key to testing the simplistic PRE model.
By 2003, 12 PRE bursters in 12 globular clusters (GCs) had distances determined from RR Lyrae stars residing in the GCs; incorporating the distances with respective peak bolometric fluxes derived from X-ray observations of PRE bursts, \citet{Kuulkers03} measured 12 observed Eddington luminosities, 9 of which are consistent with the theoretical value by \citet{Lewin93}.
The independent check by \citet{Kuulkers03} largely confirms the validity of the simplistic PRE model. The confirmation, however, is not conclusive due to the modest number of PRE bursters residing in GCs. Besides, this GC-only sample of PRE bursters might lead to some systematic offsets of the measured Eddington luminosities, as a result of potential systematic offsets of distances determined with RR Lyrae stars or X-ray flux decay that becomes more prominent in the dense regions of GCs.

Alternatively, geometric parallaxes of PRE bursters provided by the Gaia mission \citep{Gaia-Collaboration16} can also be used to test the simplistic PRE model.
This effort normally requires determination of Gaia parallax zero point $\pi_0$ for each PRE burster, as $\pi_0$ is found to be offset from 0 \citep{Lindegren18,Lindegren21}.
Recently, \citet{Arnason21} (A21) has identified Gaia Data Release 2 (DR2) counterparts for 10 PRE bursters with Gaia parallaxes. After applying the Gaia DR2 global $\pi_0$ \citep{Lindegren18}, A21 converted these parallaxes into distances, incorporating the prior information described in \citet{Bailer-Jones18}. A21 noted that the inferred distances of the 10 PRE bursters are systematically smaller than the nominal PRE distances estimated before 2008 (this discrepancy is relieved with the latest nominal PRE distances by \citealp{Galloway20}, see the discussion in Section~\ref{subsubsec:bayes_factor_about_X}).  
However, due to the predominance of low-significance parallax constraints, this discrepancy is strongly dependent on the underlying Galactic distribution of PRE bursters (as is shown in Figure~3 of A21), which is far from well constrained. 
Moreover, adopting the Gaia DR2 global $\pi_0$, instead of one individually estimated for each source, would introduce an extra systematic error.

This work furthers the effort of A21 to test the simplistic PRE model with Gaia parallaxes of PRE bursters, making use of the latest Gaia Early Data Release 3 (EDR3, \citealp{Brown20}). 
Unlike A21, this work only focuses on the PRE bursters with relatively significant Gaia parallaxes; we apply a locally-determined $\pi_0$ for each PRE burster, and probe the simplistic PRE model as well as the composition of nuclear fuel.

In addition to the above-mentioned motivation, Gaia astrometry of PRE bursters (hereafter simplified as bursters, when unambiguous) can provide more than geometric parallaxes (and hence distances). A Gaia counterpart of a burster also brings a reference position precise to sub-mas level, and possibly a detected proper motion as well.
The latter can be combined with the parallax-based or model-dependent distance estimates to yield the burster's transverse space velocity (the transverse velocity with respect to its Galactic neighbourhood).
The space velocity distribution of PRE bursters (or LMXBs) is important for testing binary evolution theories (see \citealp{Tauris17} as an analogy).
Both reference position and proper motion can be used to confirm or rule out candidate counterparts at other wavelengths, especially when the immediate neighbourhood of a burster is crowded on a $\sim$1'' scale.

In this paper, all uncertainties are quoted at the $1\,\sigma$ confidence level unless otherwise stated; all sky positions are J2000 positions.
As we mainly deal with Gaia photometric passbands \citep{Jordi10} in this work, a conventional optical passband is referred to in the form of $\mathrm{Y}^*$, while its magnitude is denoted by $Y^*$; by contrast, a Gaia passband is referred to in the form of Y band, and its magnitude is denoted by $m_\mathrm{Y}$.

\section{Gaia counterparts of PRE bursters with detected parallaxes}
\label{sec:identification}
To be able to effectively refine nominal PRE distances, constrain the composition of nuclear fuel on the NS surface and test the simplistic PRE model, we only search for Gaia counterparts with detected parallaxes $\pi_1$ ($>3\,\sigma$, a criterion we apply throughout this paper). This cutoff leads to a smaller sample of PRE bursters compared to A21, which included many marginal parallax detections. 
Our search is based on the positions of 115 PRE bursters compiled in Table~1 of \citet{Galloway20}. 
For each PRE burster, we found its closest Gaia EDR3 source within 10\farcs0 using {\tt TOPCAT}\footnote{\url{http://www.star.bris.ac.uk/~mbt/topcat/}}. 
From the resultant 110 candidates, we shortlisted 16 Gaia counterpart candidates with detected $\pi_1$. 
Among the 16 Gaia counterpart candidates, the Gaia counterparts for \cyg, \cen\ and \uua\ have been recognized by VizieR \citep{Ochsenbein00} in an automatic manner, simply based on the $\approx1$\,mas angular distance (at the reference epoch year 2015.5 disregarding proper motion) that is comparable to the Gaia positional uncertainties (see Table~\ref{tab:Gaia_counterparts}); the Gaia counterpart for XB~2129$+$47 has been identified by A21.

To identify Gaia counterparts of PRE bursters from the remaining 12 candidates, we adopted a more complicated cross-match criterion,
which requires the identification of an optical counterpart (either confirmed or potential) for the PRE burster. For an optical source $\mathcal{S}$ that is a confirmed or potential counterpart, we consider it associated with the Gaia source $\mathcal{G}$ if all the following conditions are met:
\begin{enumerate}[label=(\roman*)]
    \item $\mathcal{S}$ is sufficiently bright ($3\leq G^*\leq 21$\footnote{\url{https://www.cosmos.esa.int/web/gaia/earlydr3}}) for Gaia detection; the conventional apparent magnitude measured closest to $\mathrm{G}^*$ band is looked at if $G^*$ is not available (as compiled in Table~\ref{tab:Gaia_counterparts});
    \item $\mathcal{G}$ falls into the 1-$\sigma_\mathcal{S}$-radius circle around $\mathcal{S}$ (where the position of a confirmed radio or infrared counterpart of $\mathcal{S}$ would be adopted if it is more precise than the optical position);
    \item $\mathcal{G}$ is the only Gaia source within the 5-$\sigma_\mathcal{S}$-radius circle around $\mathcal{S}$.
\end{enumerate}
Here, to account for the effect of proper motion (and the smaller contribution of parallax), 
the position of the candidate has been extrapolated to 
the epoch at which the position of $\mathcal{S}$ was measured, with the astrometric parameters of $\mathcal{G}$ using the ``predictor mode'' of {\tt pmpar} (available at \url{https://github.com/walterfb/pmpar}). This extrapolation allows the evaluation of the same-epoch $\Delta_\mathcal{S-G}$, the separation between $\mathcal{S}$ and $\mathcal{G}$, which is free from proper motion and parallax effects. 

In principle, the apparent magnitude of $\mathcal{G}$ should match that of the optical counterpart of $\mathcal{S}$. However, making such a comparison is complicated by {\bf 1)} the wide photometric passbands designed for the in-flight Gaia \citep{Jordi10} and {\bf 2)} magnitude variability of stars. 

On top of the 4 Gaia counterparts of PRE bursters (\cen, \cyg, \uua\ and \xb) already recognized by VizieR and A21, we identified one Gaia counterpart with detected $\pi_1$ (from the  remaining 12 objects), which was NP~Ser (see Table~\ref{tab:Gaia_counterparts}). However, as we will discuss in Section~\ref{subsec:NpSer_neq_Gx17}, NP~Ser is not the optical counterpart of \gx, meaning that our final sample consists of 5 Gaia sources with detected $\pi_1$, of which 4 are PRE bursters.

Hereafter, the 5 Gaia sources are sometimes referred to as ``targets''.
The (Gaia) B band covers the conventional $\mathrm{G}^*$, $\mathrm{V}^*$, $\mathrm{B}^*$ bands and part of the $\mathrm{R}^*$ band. According to Table~\ref{tab:Gaia_counterparts}, the magnitude $V^*$ of NP~Ser measured at conventional $\mathrm{V}^*$ band generally agrees with $m_\mathrm{B}$ of its Gaia counterpart.
The 5 astrometric parameters of the Gaia counterparts for the 4 PRE bursters and NP~Ser are summarized in Table~\ref{tab:before_calibration}.

\begin{table*}
\caption{Gaia EDR3 counterparts with detected parallaxes $\pi_1$ for 4 PRE bursters and NP~Ser}
\label{tab:Gaia_counterparts}
\centering
\begin{tabular}{@{}cccccccc@{}}
\hline\hline
source name & $m^*$ $^{a_1}$ & $\sigma_\mathcal{S}$ $^{a_2}$ & Gaia EDR3 & $m_\mathrm{B}$ $^{a_3}$ & $\Delta_\mathcal{S-G}$ $^{a_4}$ & $R_1$ $^{a_5}$ & associated\\
& (mag) & ('') & counterpart & (mag) & ('') & ('') & by? \\
\hline%
NP~Ser & $V^*\!=\!17.42 ^{e_1}$ & 0.5$^{e_2}$ & 4146621775597340544 & 17.72 & 0.3 & 12 & this work\\
\cyg\ & $-$ & $-$ & 1952859683185470208 & $-$ & $-$ & $-$ & VizieR$^i$\\
\cen\ & $-$ & $-$ & 6205715168442046592 & $-$ & $-$ & $-$ & VizieR$^i$\\
\uua\ & $-$ & $-$ & 5310395631798303104 & $-$ & $-$ & $-$ & VizieR$^i$\\
\xb\ & $-$ & $-$ & 1978241050130301312 & $-$ & $-$ & $-$ & A21$^k$\\
\hline\hline
\end{tabular}

\tabnote{$^{a_1}$Apparent magnitude at a conventional passband;
$^{a_2}$positional uncertainty (adding in quadrature uncertainties in both directions) of the source (see Section~\ref{sec:identification}); 
$^{a_3}$apparent magnitude at the Gaia B band \citep{Jordi10}, which covers the conventional $\mathrm{G^*}$, $\mathrm{V^*}$, $\mathrm{B^*}$ bands and part of the $\mathrm{R^*}$ band;
$^{a_4}$angular separation between the source and its Gaia counterpart, where $\alpha_\mathcal{G}$ and $\delta_\mathcal{G}$ were extrapolated to the respective epoch of $\alpha_\mathcal{S}$ (and $\delta_\mathcal{S}$) using the astrometric parameters in Table~\ref{tab:before_calibration} (see Section~\ref{sec:identification} for more explanation); 
$^{a_5}$maximum search radius that contains only one Gaia source.}

\tabnote{$^{e_1}$\citet{Deutsch96};
$^{e_2}$\citet{Deutsch99}, where $\alpha_\mathcal{S}=18^{\rm h}16^{\rm m}01\fs$380, $\delta_\mathcal{S}=-$14\degr02'11\farcs34 out of R-band CCD observation, with uncertainty radius of 0\farcs5 at 90\% confidence, measured at MJD~50674.}

\tabnote{
$^i$\citet{Ochsenbein00}; $^k$\citet{Arnason21}.}
\end{table*}

\begin{table*}
\caption{Five astrometric parameters from Gaia EDR3 counterparts at year 2016.0 prior to calibration of Gaia parallaxes}
\label{tab:before_calibration}
\centering
\begin{tabular}{@{}cccccc@{}}
\hline\hline
source name & $\alpha_\mathcal{G} \pm \sigma_\alpha$ & $\delta_\mathcal{G}$ & $\pi_1$ & $\mu_\alpha \equiv \dot{\alpha} \cos\delta$ & $\mu_\delta$ \\
& ($\sigma_\alpha$ in mas) &  & (mas) & (\maspy) & (\maspy)\\
\hline%
NP~Ser & $18^{\rm h}16^{\rm m}01\fs 392863 \pm 0.07$ & $-14\degr02'11\farcs 75580(6)$ & 0.69(8) & 3.35(8) & -6.97(6) \\
\cyg\ & $21^{\rm h}44^{\rm m}41\fs 152001 \pm 0.01$ & $38\degr19'17\farcs 06138(1)$ & 0.068(19) & -1.79(2) & -0.32(2) \\
\cen\ & $14^{\rm h}58^{\rm m}21\fs 93584 \pm 0.1$ & $-31\degr40'08\farcs 40635(9)$ & 0.53(13) & 0.84(15) & -55.68(13) \\
\uua\ & $09^{\rm h}20^{\rm m}26\fs 471033 \pm 0.05$ & $-55\degr12'24\farcs 47694(6)$ & 0.24(6) & -5.77(7) & 2.24(8) \\
\xb\ & $21^{\rm h}31^{\rm m}26\fs 209631 \pm 0.06$ & $47\degr17'24\farcs 44432(7)$ & 0.50(8) & -2.34(8) & -4.23(8) \\
\hline\hline
\end{tabular}
\end{table*}

\section{Calibration of Gaia parallaxes}
\label{sec:parallax_calibration}
Before being applied for scientific purposes, each of the Gaia parallaxes $\pi_1$ needs to be calibrated by determining its (Gaia) parallax zero point $\pi_0$, as $\pi_0$ has been shown to be systematically offset from zero \citep[with an all-sky median of $-0.02$\,mas for EDR3;][]{Lindegren21}.
Previous studies have revealed the dependence of $\pi_0$ on sky position, apparent magnitude and color \citep{Lindegren18,Chen18,Huang21}. Furthermore, \citet{Lindegren21} proposed that the position dependence of $\pi_0$ can be mostly attributed to the evolution of $\pi_0$ with respect to ecliptic latitude $\beta$, and derived an empirical global solution for five-parameter (see \citealp{Lindegren21a} for explanation) Gaia EDR3 sources and another such solution for six-parameter sources.
While this approach is convenient to use, the empirical solutions do not provide an estimate of the uncertainty of $\pi_0$, and are only considered indicative \citep{Lindegren21}. 

In this work, we attempted a new pathway of generic $\pi_0$ determination. Unlike the global empirical solutions, the new $\pi_0$ determination technique offers locally acquired $\pi_0$, and provides the uncertainty of the determined $\pi_0$.
We estimated $\pi_0$ for the 5 targets using a subset of the {\tt agn\_cross\_id} table publicized along with EDR3 (Klioner et al. in preparation).
We selected 1,592,629 quasar-like sources (out of the 1,614,173 sources in the {\tt agn\_cross\_id} table), which {\bf a)} have $m_\mathrm{B-R}$ (denoted as ``{\tt bp\_rp}'' in EDR3) values and {\bf b)} show no detected ($>3\,\sigma$ level) proper motion in either direction. Hereafter, we refer to this subset of 1,592,629 sources as the quasar catalog, and its sources as background quasars or quasars (when unambiguous).

On the sky, spatial patterns of $\pi_0$ on angular scales of up to $\sim10$\degr\ are seen in Figure~2 of  \citet{Lindegren21}.
We attempted different radius cuts (around each of the 5 targets) of up to 40\degr, but found no clear evidence showing quasar parallaxes representative of the target $\pi_0$ beyond a 10\degr\ radius cut.
In most cases, a 10\degr\ radius cut provides a sufficiently large sample of quasars for $\pi_0$ estimation. Hence, we only present analysis of quasars within 10\degr\ around each target.

\subsection{A new pathway to parallax zero points}
\label{subsec:s_pi0}

Our $\pi_0$ determination is solely based on the quasar catalog (of 1,592,629 sources).
The vital step of (Gaia) parallax calibration (or $\pi_0$ determination) is to select appropriate quasars in the same sky region with similar G-band apparent magnitudes $m_\mathrm{G}$ and colors $m_\mathrm{B-R}$.
We implemented this selection for each of the 5 targets by applying 4 filters to the quasar catalog: the angular-distance filter, the $\beta$ filter, the $m_\mathrm{G}$ filter and the $m_\mathrm{B-R}$ filter. 
Once appropriate background quasars are chosen for each target, one can calculate $\pi_0$ as the weighted mean parallax of this ``sub-sample'', and the formal uncertainty of $\pi_0$
\begin{equation}
    \tilde{\sigma}_{\pi_0}=\left({\sum\limits_{i}\frac{1}{{\sigma_i}^2}}\right)^{-\frac{1}{2}}
	\label{eq:parallax_zero_point_uncertainty}
\end{equation}
(Equation~4.19 in \citealp{Bevington03}), where $\sigma_i$ is the (Gaia) parallax error of each background quasar in the sub-sample. 

We parameterized {\bf 1)} the angular-distance filter using the search radius $r$ around the target, {\bf 2)} the $\beta$ filter using $\Delta \sin{\beta}$, the half width of the $\sin{\beta}$ filter centred about $\sin{\beta^\mathrm{target}}$, {\bf 3)} the $m_\mathrm{G}$ filter using $\underline{\Delta} m_\mathrm{G}$, which stands for the relative half width (or $\Delta m_\mathrm{G}/m_\mathrm{G}^\mathrm{target}$) of the $m_\mathrm{G}$ filter around $m_\mathrm{G}^\mathrm{target}$, and {\bf 4)} the $m_\mathrm{B-R}$ filter using $\Delta m_\mathrm{B-R}$, the half width of the $m_\mathrm{B-R}$ filter centred about $m_\mathrm{B-R}^\mathrm{target}$.

\begin{figure*}
    \centering
	\includegraphics[width=14cm]{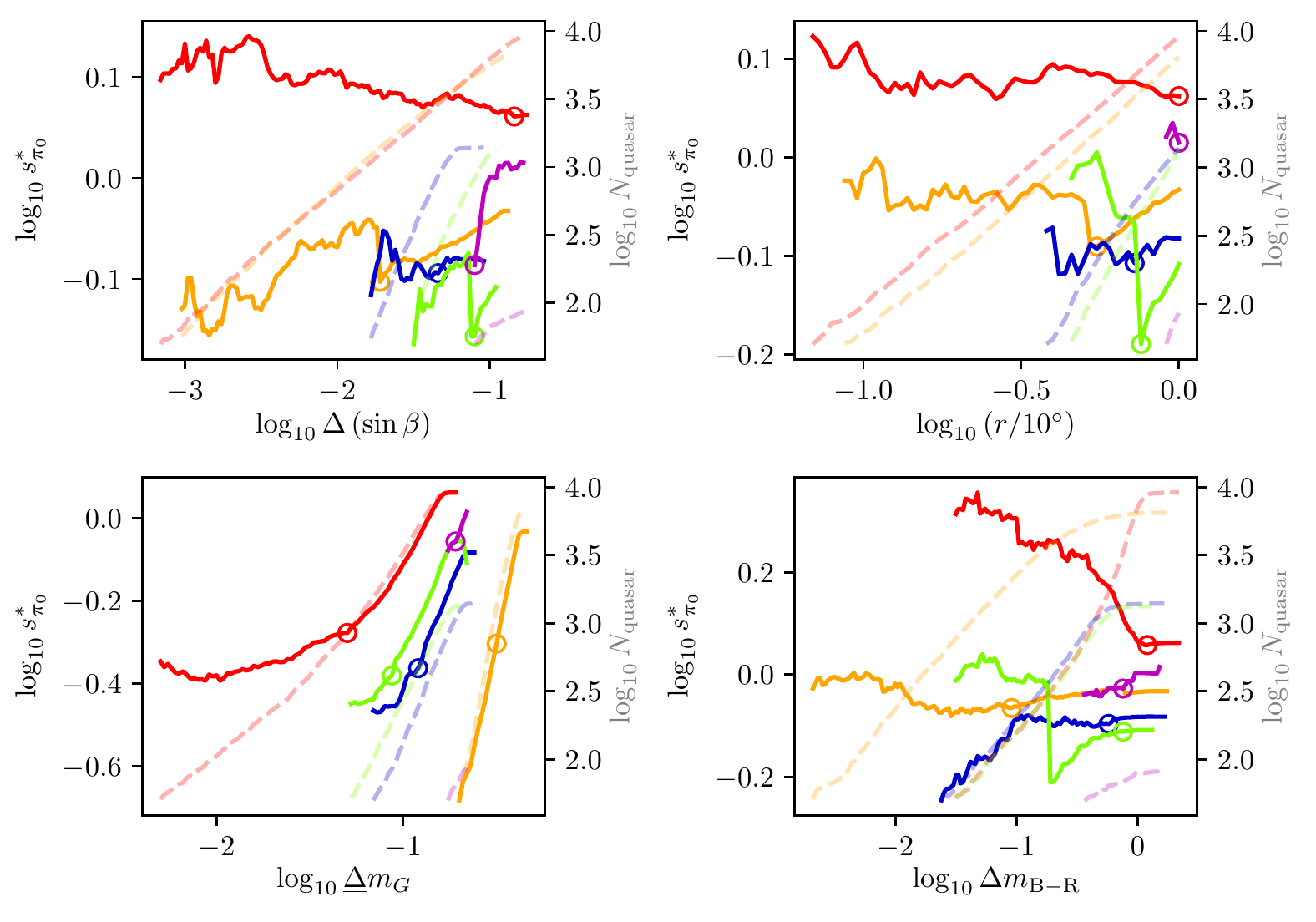}
	\includegraphics[width=13cm]{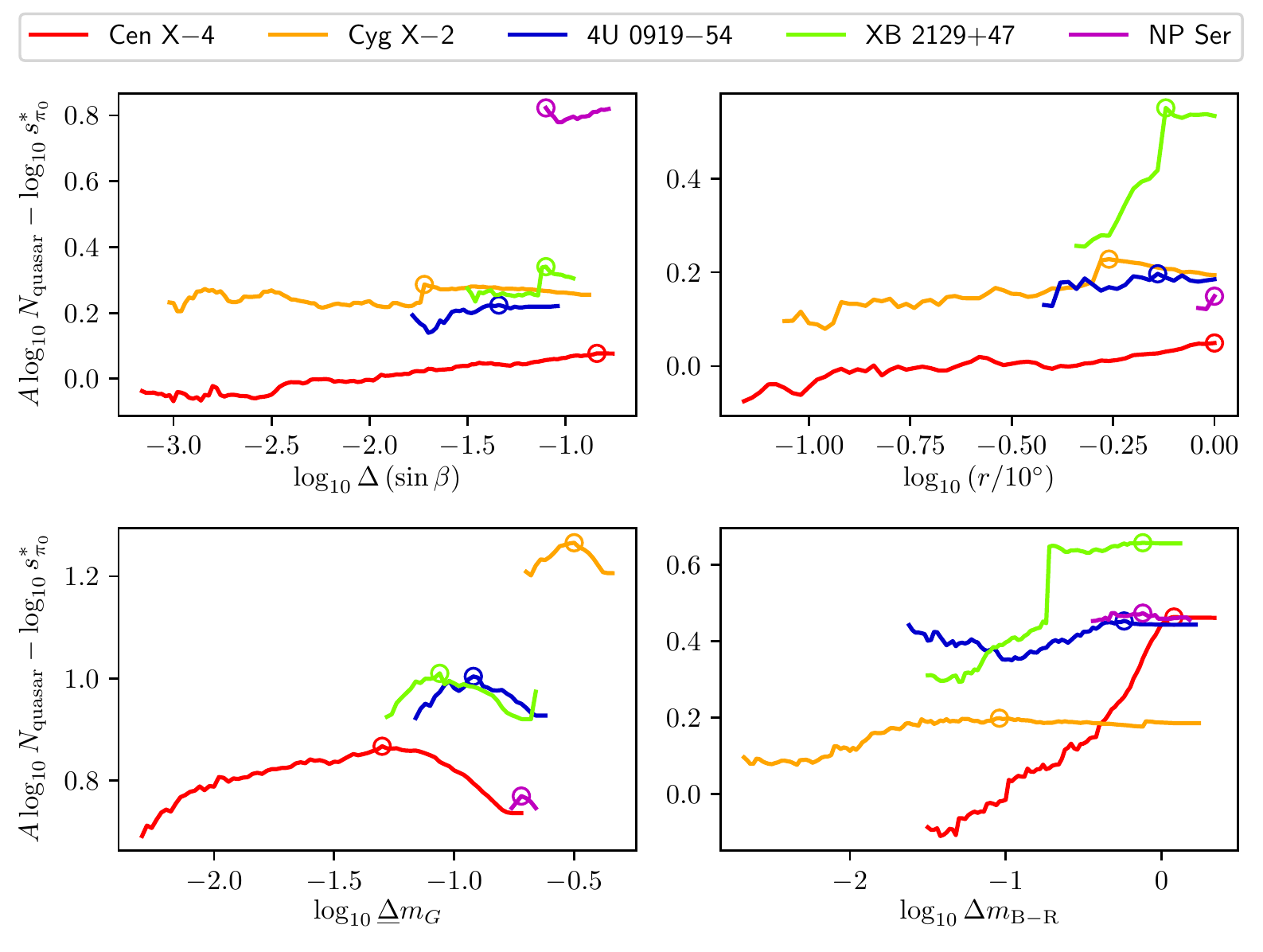}
    \caption{{\bf Top 4-panel block:} The solid lines present the marginalized relations between $s_{\pi_0}^{*}$ of background quasars and 4 filter parameters: $\Delta \sin{\beta}$ ($\beta$ denotes ecliptic latitude), search radius $r$ around the target, $\underline{\Delta} m_\mathrm{G}$ and $\Delta m_\mathrm{B-R}$.
    Here, $s_{\pi_0}^{*}=s_{\pi_0}/s_{\pi_0}^{\rm glb}$, where $s_{\pi_0}$ and $s_{\pi_0}^{\rm glb}$ represent the weighted standard deviation of quasar parallaxes and its global value (of the 1592629 quasars in the quasar catalog), respectively; 
    $\Delta m_\mathrm{B-R}$|$\Delta \sin{\beta}$ stands for the half width of the $m_\mathrm{B-R}$|$\sin{\beta}$ filter centred around the $m_\mathrm{B-R}^\mathrm{target}$|$\sin{\beta^\mathrm{target}}$ (used to pick like-$m_\mathrm{B-R}$|like-$\beta$ background quasars); $\underline{\Delta} m_\mathrm{G}$ defines the half relative width of the $m_\mathrm{G}$ filter around $m_\mathrm{G}^\mathrm{target}$ used to select like-$m_\mathrm{G}$ background quasars.
    The dashed curves lay out the marginalized relations between $N_\mathrm{quasar}$ and the 4 filter parameters, where $N_\mathrm{quasar}$ stands for number of remaining background quasars after being filtered. 
    To avoid small-sample effect, the calculations start from $N_{\rm quasar}=50$, or $N_{\rm quasar}^{\rm start}=50$.
    The dependence of $\pi_0$ on a variable ($r$, $\beta$, $m_\mathrm{G}$ or $m_\mathrm{B-R}$) is suggested if $s_{\pi_0}^{*}$ grows with a larger filter (such as the $m_\mathrm{G}$ dependence), and vice versa.
    {\bf Bottom 4-panel block:} 
    The marginalized relation between the index $w=A \log_{10}N_{\rm quasar}-\log_{10}s_{\pi_0}^{*}$ and each filter parameter, where $A=(\log_{10}s_{\pi_0}^{\rm *,max}-\log_{10}s_{\pi_0}^{\rm *,min})/(\log_{10}N_{\rm quasar}^{\rm max}-\log_{10}N_{\rm quasar}^{\rm start})$ is a scaling factor. The maximum of $w$ is chosen (circled out in both bottom and top blocks) as the ``optimal'' filter parameter.
    }
    \label{fig:S__DM_G__relation}
\end{figure*}

Hence, the problem of $\pi_0$ determination using background quasars can be reduced to searching for optimal filter parameters (for each target), i.e. $r$, $\Delta \sin{\beta}$, $\underline{\Delta} m_\mathrm{G}$ and $\Delta m_\mathrm{B-R}$. 
For this search, we investigated the marginalized relation between $s_{\pi_0}^{*}$ and each of the 4 parameters (see the top 4-panel block of Figure~\ref{fig:S__DM_G__relation}), where $s_{\pi_0}^{*}=s_{\pi_0}/s_{\pi_0}^\mathrm{glb}$ stands for the weighted standard deviation of quasar parallaxes divided by its global value of 0.32\,mas. 
To avoid small-sample fluctuations, Figure~\ref{fig:S__DM_G__relation} starts at $N_\mathrm{quasar}=50$, or $N_\mathrm{quasar}^\mathrm{start}=50$, which can also prevent overly small-sized sub-samples.
To start with, we first introduce the $s_{\pi_0}^{*}$-to-$\underline{\Delta} m_\mathrm{G}$ relation and explain its implications. Relations between $s_{\pi_0}^{*}$ and other 3 parameters can be interpreted in the same way.

Unlike $\tilde{\sigma}_{\pi_0}$, if a sufficiently large ($\gtrsim50$) background-quasar sample had no $m_\mathrm{G}$ dependence (as opposed to \citealp{Lindegren18,Lindegren21}), $\log_{10} s_{\pi_0}^{*}$ would be expected to be largely independent to $\underline{\Delta} m_\mathrm{G}$, and to fluctuate around the global value 0.
This is clearly not the case for all of the 5 targets. From Figure~\ref{fig:S__DM_G__relation}, we found an obvious upward trend of $s_{\pi_0}^{*}$ with growing $\underline{\Delta} m_\mathrm{G}$. This upward trend shows that background quasars with similar $m_\mathrm{G}$ around $m_\mathrm{G}^\mathrm{target}$ tend to have similar parallaxes, which supports the $m_\mathrm{G}$-dependence of $\pi_0$ stated in \citet{Lindegren18,Lindegren21}.

Using the same interpretation, we also investigated marginalized relations of $s_{\pi_0}^{*}$ with respect to each of the other 3 filter parameters. As can be seen in Figure~\ref{fig:S__DM_G__relation}, most fields show a (weak) color dependence of $\pi_0$. For the two sky-position-related filter parameters, $\beta$ dependence of $\pi_0$ is clearly found in the target fields of \cyg\ and NP~Ser, whereas the $r$ dependence of $\pi_0$ is hardly noticeable in any of the 5 target fields. This contrast reinforces the belief that the $\beta$ dependence of $\pi_0$ contributes considerably to the position dependence of $\pi_0$ \citep{Lindegren21}. Additionally, $\log_{10} s_{\pi_0}^{*}$ converges to $0.0\pm0.1$ at $r=10\degr$ in all target fields, which roughly agrees with our assumption that the $r$-dependence of $\pi_0$ becomes negligible beyond 10\degr\ search radius; in other words, quasar parallaxes at $r>10\degr$ are almost not representative of the target $\pi_0$.

Knowing the way to interpret the relation between $s_{\pi_0}^{*}$ and each of the filter parameters, we can proceed to choose the optimal filter parameters. In choosing filter parameters, one wants the filtered sub-sample to have a consistently low level of $s_{\pi_0}^{*}$; meanwhile, one prefers a filtered sub-sample as large as possible to achieve reliable and precise $\pi_0$. To meet both demands is difficult when $s_{\pi_0}^{*}$ rises with a larger filter (as is the case for the $m_\mathrm{G}$ filter). In this regard, we created an index $w=A \log_{10} N_\mathrm{quasar} - \log_{10} s_{\pi_0}^{*}$, where $A=(\log_{10}s_{\pi_0}^{\rm *,max}-\log_{10}s_{\pi_0}^{\rm *,min})/(\log_{10}N_{\rm quasar}^{\rm max}-\log_{10}N_{\rm quasar}^{\rm start})$ is a scaling factor. Instead of the minimum of $s_{\pi_0}^{*}$ or the maximum of $N_{\rm quasar}$, we adopted the filter parameter at the peak of the index $w$ (see the bottom 4-panel block of Figure~\ref{fig:S__DM_G__relation}), so that an optimized trade-off between $s_{\pi_0}^{*}$ and $N_{\rm quasar}$ can be reached. Using this consistent standard, we reached the parameter of each filter for each target (see Table~\ref{tab:zero_parallax_points}).

After applying the 4 filters, we acquired a quasar sub-sample of $N_\mathrm{quasar}$ and $s_{\pi_0}^{*}$ for each target, and obtained $\pi_0$ of the target from this sub-sample (see Table~\ref{tab:zero_parallax_points}).
Due to the relative paucity of quasars identified at low Galactic latitudes $b$ (for the reasons discussed in  \citealp{Lindegren21}), NP~Ser, \uua\ and \xb\ (all of which have $|b|<4$\degr) receive relatively low values for $N_\mathrm{quasar}$ compared to \cen\ ($b=24$\degr), which partly lead to the relatively large $\tilde{\sigma}_{\pi_0}$. 
Nonetheless, we have shown that $\pi_0$ for targets at $|b|<6$\degr\ can be determined with nearby (on the sky) quasars.
On the other hand, \cyg, despite being located at $b=-11$\degr, has the smallest $N_\mathrm{quasar}$ and the largest $\tilde{\sigma}_{\pi_0}$ among the 5 targets.  This is mainly due to the relative rarity of bright quasars (\cyg\ has $m_\mathrm{G}=14.7$\,mag).

The negative $\log_{10} s_{\pi_0}^{*}$ of the quasar sub-sample for each target indicates that the sub-sample of (closely located, like-$\beta$, like-$m_\mathrm{G}$ and like-$m_\mathrm{B-R}$) quasars are representative of the target source.
Despite this good representativeness, the $\tilde{\sigma}_{\pi_0}$ estimated with Equation~\ref{eq:parallax_zero_point_uncertainty} is still an under-estimate for the uncertainty of $\pi_0$, as the sub-sample has a spread in each parameter and does not represent the target perfectly. We estimated the weighted average $m_\mathrm{G}$, $m_\mathrm{B-R}$ and $\sin{\beta}$ of the quasar sub-sample (where the weighting is $1/{\sigma_i}^2$, see Equation~\ref{eq:parallax_zero_point_uncertainty} for the definition of $\sigma_i$), which are presented in Table~\ref{tab:zero_parallax_points} as $\overline{m_\mathrm{G}}$, $\overline{m_\mathrm{B-R}}$ and $\overline{\sin{\beta}}$, respectively.
According to Table~\ref{tab:zero_parallax_points}, there are residual offsets of $\sin{\beta}$, $m_\mathrm{G}$ and $m_\mathrm{B-R}$ between the target and the average level of the sub-sample, which would result in extra systematic error for the $\pi_0$ estimates.
In order to estimate this systematic error of $\pi_0$, we calculated the empirical parallax zero points for the targets (see Table~\ref{tab:zero_parallax_points}) using {\tt zero\_point.zpt} (\url{https://gitlab.com/icc-ub/public/gaiadr3_zeropoint}, \citealp{Lindegren21}), noted as $\pi_0^\mathrm{emp}$.
In the same way, we also estimated the empirical parallax zero points for each quasar of the sub-sample, then derived the weighted average empirical parallax zero point of the sub-sample, noted as $\overline{\pi_0^\mathrm{emp}}$ (see Table~\ref{tab:zero_parallax_points}). The difference between $\pi_0^\mathrm{emp}$ and $\overline{\pi_0^\mathrm{emp}}$ is taken as an estimate for the systematic error of $\pi_0$, which is added in quadrature to the $\tilde{\sigma}_{\pi_0}$ calculated with Equation~\ref{eq:parallax_zero_point_uncertainty}.

No uncertainties are yet available for $\pi_0^\mathrm{emp}$. Regardless, all of our $\pi_0$ are consistent with the empirical counterparts at the $2\,\sigma$ confidence level.
We note that $\pi_0^\mathrm{emp}$ is only used to estimate the systematic error of $\pi_0$, and is not adopted in the discussions that follow.
The calibrated parallaxes $\pi_1-\pi_0$ are summarized in Table~\ref{tab:zero_parallax_points}.

Despite the overall consistency between $\pi_0$ and $\pi_0^\mathrm{emp}$, it is noteworthy that all $\pi_0$ are larger than the $\pi_0^\mathrm{emp}$ counterparts, which can be a coincidence of small-number statistics. 
Alternatively, knowing that most of the targets are situated at low $b$, it might indicate a systematic offset between quasar-sub-sample-based $\pi_0$ and $\pi_0^\mathrm{emp}$ at low $b$.  
Interestingly, a recent parallax zero-point study based on $\sim$110,000 W Ursae Majoris variables showed that the W-Ursae-Majoris-variable-based $\pi_0$ are systematically larger than the $\pi_0^\mathrm{emp}$ counterparts at low $b$ (see Figure~3 of \citealp{Ren21}). 
Hence, a future quasar-sub-sample-based study involving a large sample of targets (and other parallax zero-point studies by different approaches) will be essential for probing $\pi_0^\mathrm{emp}$ at low $b$.
More specifically, if quasar-sub-sample-based $\pi_0$ at low $b$ are confirmed to be systematically larger than the $\pi_0^\mathrm{emp}$ counterparts, then it is likely that $\pi_0^\mathrm{emp}$ is systematically under-estimated at low $b$.

\begin{table*}
\caption{The information of the 4 filters (including search radius $r$, $\beta$ filter, $m_\mathrm{G}$ filter and $m_\mathrm{B-R}$ filter, see Section~\ref{subsec:s_pi0} for explanation) and the parallax zero point $\pi_0$ calculated from the respective sub-sample of background quasars after applying the 4 filters. $\pi_1$ and $\pi_1-\pi_0$ stand for uncalibrated parallaxes and calibrated parallaxes, respectively. 
$\overline{m_\mathrm{G}}$, $\overline{m_\mathrm{B-R}}$ and $\overline{\sin{\beta}}$ represent the respective weighted average value of the three filter parameters (see Section~\ref{subsec:s_pi0}).
$N_\mathrm{quasar}$ and $s_\mathrm{\pi_0}^{*}$ (defined in Section~\ref{subsec:s_pi0}) are reported for the filtered quasar sub-sample in each target field. 
The empirical parallax zero-point solutions for the targets calculated with {\tt zero\_point.zpt} (\url{https://gitlab.com/icc-ub/public/gaiadr3_zeropoint}) are provided as $\pi_0^\mathrm{emp}$. The weighted average empirical parallax zero-points of the sub-samples (see Section~\ref{subsec:s_pi0}) are presented as $\overline{\pi_0^\mathrm{emp}}$.
}
\label{tab:zero_parallax_points}
\centering
\begin{tabular}{@{}ccccccccccc@{}}
\hline\hline
field & $r$ & $\sin{\beta^\mathrm{target}}$ & $\Delta \sin{\beta}$ & $m_\mathrm{G}^\mathrm{target}$ & $\underline{\Delta} m_\mathrm{G}$ & $m_\mathrm{B-R}^\mathrm{target}$ & $\Delta m_\mathrm{B-R}$ & $\overline{m_\mathrm{G}}$ & $\overline{m_\mathrm{B-R}}$ & $\overline{\sin{\beta}}$\\
& (\degr) &  & & (mag) &  & (mag) & (mag) & & &\\
\hline%
\cen\ & 10 & $-0.24$ & 0.14 & 17.85 & 5.0\% & 1.59 & 1.20 & 18.00 & 0.74 & -0.25 \\
\cyg\ & 5.5 & 0.74 & 0.02 & 14.70 & 31.6\% & 0.71 & 0.09 & 18.19 & 0.74 & 0.74\\
\uua\ & 7.2 & $-0.90$ & 0.05 & 17.15 & 12.0\% & 1.19 & 0.58 & 18.23 & 0.94 & -0.93\\
\xb\ & 7.6 & 0.84 & 0.08 & 17.58 & 8.7\% & 1.29 & 0.76 & 18.49 & 1.01 & 0.79\\
NP~Ser & 10 & 0.16 & 0.08 & 17.01 & 19.1\% & 1.51 & 0.76 & 17.61 & 1.57 & 0.16\\
\hline\hline
\end{tabular}

\vspace{0.3cm}

\begin{tabular}{@{}cccccccc@{}}
\hline\hline
field & $N_\mathrm{quasar}$  &  $\log_{10}s_{\pi_0}^{*}$ &  $\pi_0^\mathrm{emp}$ & $\overline{\pi_0^\mathrm{emp}}$ & $\pi_0$ & $\pi_1$ & $\pi_1-\pi_0$\\
& & & (mas) & (mas) & (mas) & (mas) \\
\hline%
\cen\ & 772 & $-0.27$ & $-0.027$ & $-0.026$ & $-0.022\pm0.006\pm0.002$ & $0.53\pm0.13$ & $0.55\pm0.13$ \\
\cyg\ & 32 & $-0.29$ & $-0.030$ & $-0.020$ & $0.019\pm0.024\pm0.010$ &  $0.068\pm0.019$ & $0.050\pm0.032$\\
\uua\ & 91 & $-0.33$ & $-0.028$ & $-0.027$ & $-0.009\pm0.013\pm0.001$ &  $0.24\pm0.06$ & $0.25\pm0.06$\\
\xb\ & 63 & $-0.37$ & $-0.028$ & $-0.019$ & $0.004\pm0.018\pm0.009$ &  $0.50\pm0.08$ & $0.50\pm0.08$\\
NP~Ser & 40 & $-0.17$ & $-0.032$ & $-0.026$ & $-0.008\pm0.033\pm0.006$ & $0.69\pm0.08$ & $0.70\pm0.09$\\
\hline\hline
\end{tabular}
\end{table*}

\section{Discussion}
\label{sec:discussions}
In this section, we discuss the implications of the 5 calibrated parallaxes $\pi_1-\pi_0$ provided in Table~\ref{tab:zero_parallax_points}. 

\subsection{NP~Ser and \gx\ revisited}
\label{subsec:NpSer_neq_Gx17}
\gx\ is one of the brightest X-ray sources on the sky, from which 43 PRE bursts have been recorded (see Table~1 of \citealp{Galloway20} and references therein); its distance was estimated to be 7.3--12.6\,kpc by treating its PRE bursts as standard candles (as explained in Section~\ref{sec:intro}), where the distance uncertainty is dominated by the uncertain composition of the nuclear fuel \citep{Galloway20}. 
The optical source NP~Ser, $\approx1\farcs0$ away from \gx, was initially considered the optical counterpart of \gx\ \citep{Tarenghi72}; but the association has been ruled out, based on its sky-position offset from \gx\ \citep{Deutsch99} and its lack of optical variability (as expected for the optical counterpart of \gx) \citep[e.g.][]{Davidsen76,Margon78}.
Despite this, NP~Ser was initially treated as a potential counterpart and the Gaia properties of NP~Ser were analysed.
Regardless of the non-association of NP~Ser and \gx, the complex sky region around \gx\ for optical/infrared observations (see Figure~2 of \citealp{Deutsch99} for example) merits high-precision Gaia astrometry of NP~Ser, which will facilitate future data analysis of optical/infrared observations of \gx.

Using the astrometric parameters of NP~Ser in Tables~\ref{tab:before_calibration} and \ref{tab:zero_parallax_points}, we extrapolated the reference position of NP~Ser (from the reference epoch 2016.0\,yr, or MJD~57388) to MJD~47496, following the method described in Section~3.2 of \citet{Ding20}. 
The projected Gaia position of NP~Ser is offset from the VLA position of \gx\ (see Table~2 of \citealp{Deutsch99}) by 0\farcs$95\pm0$\farcs07 at MJD~47496.
Considering the distances to \gx\ and NP~Ser, nuclear fuel with cosmic abundances (73\% hydrogen) corresponds to the smallest PRE distance of $8.5\pm1.2$\,kpc for \gx\ \citep{Galloway20};
in comparison, the calibrated Gaia parallax of NP~Ser is $0.70\pm0.09$\,mas, corresponding to a distance of $1.44^{+0.21}_{-0.16}$\,kpc, which establishes NP~Ser as a foreground source of \gx.
The current sky-position offset between NP~Ser and \gx\ depends on the proper motion of \gx. Assuming \gx\ rotates around the Galactic centre in the Galactic disk with negligible peculiar velocity (with respect to the neighbourhood of \gx), the apparent proper motion of \gx\ would be $\mu_{\alpha}=-3.5$\,\maspy\ and $\mu_{\delta}=-6.7$\,\maspy\ given a distance of 8.5\,kpc (from the Earth).
Based on this indicative proper motion of \gx\ and the Gaia ephemeris of NP~Ser, NP~Ser is currently (at MJD~59394) 1\farcs0 away from \gx, and continues to move away from \gx\ on the sky; accordingly, it would become easier with time to resolve \gx\ from the foreground NP~Ser in optical/infrared observations.

\subsection{Posterior distances and compositions of nuclear fuel for \cen, \cyg\ and \uua}
\label{subsec:posterior_distances}
Among the (calibrated) parallaxes $\pi_1-\pi_0$ of the 4 PRE bursters (i.e. \cen, \cyg, \uua\ and \xb), $\pi_1-\pi_0$ for \cen, \uua\ and \xb\ are detected, whereas $\pi_1-\pi_0$ of \cyg\ is weakly constrained.
All PRE bursters except \xb\ have nominal PRE distances published that are inferred from their respective PRE bursts (see Table~\ref{tab:constrain_composition} for PRE distances and their references).
Using Bayesian inference, we can incorporate the parallax of a PRE burster with its nominal PRE distance to constrain the composition of nuclear fuel and refine the nominal PRE distance. 

In Section~\ref{sec:intro}, we have mentioned that the Eddington luminosity (measured by an observer at infinity) $L_{\mathrm{Edd},\infty}$ varies with NS mass $M_\mathrm{NS}$ and photosphere radius $R_\mathrm{P}$ (see Equation~\ref{eq:Eddington_luminosity}). In practice, the bolometric flux at the ``touchdown'' (when the photosphere drops back to the NS surface) of a PRE burst is compared to the theoretical $L_{\mathrm{Edd},\infty}$ at $R_\mathrm{P}=R_\mathrm{NS}$ (where $R_\mathrm{NS}$ stands for NS radius) to calculate the nominal PRE distance \citep{Galloway20}. Therefore, the theoretical $L_{\mathrm{Edd},\infty}$ should change with $M_\mathrm{NS}$ and $R_\mathrm{NS}$, which is, however, not taken into account in this work.
We note that, following \citet{Galloway20}, we assume $M_\mathrm{NS}=1.4\,\msun$ and $R_\mathrm{NS}=11.2$\,km \citep{Steiner18} for all PRE bursters.

\subsubsection{Bayes factor between two ends of the composition of nuclear fuel}
\label{subsubsec:bayes_factor_about_X}
As mentioned in Section~\ref{sec:intro}, the Eddington luminosity of PRE bursts depends on the composition of nuclear fuel (accreted onto the NS surface from its companion) at the time of the outburst. This composition is normally parameterized by $X$, the mass fraction of hydrogen in nuclear fuel at the time of a PRE burst, which generally ranges from 0 to $\approx73$\%, the cosmic mass fraction of hydrogen. 
As the result of the dynamic nucleosynthesis process during accretion, $X$ varies from one PRE burst to another, but is always smaller than the hydrogen mass fraction of the donor star. 
$X$ can be roughly predicted by the theoretical ignition models of X-ray bursts, and mainly depends on the local accretion rate $\dot{m}$ \citep[e.g.][]{Fujimoto81}.

Nominal PRE distances for $X=0$, denoted as $D_0$, are summarized in Table~\ref{tab:constrain_composition} along with their references.
At any given $X$, the nominal PRE distance $D_X= D_0/\sqrt{X+1}$ \citep{Lewin93}. As the fractional uncertainty of a nominal PRE distance measurement is independent of $X$, $\sigma_X=\sigma_0/\sqrt{1+X}$, where $\sigma_0$ and $\sigma_X$ represent the uncertainty of the nominal PRE distance at $X=0$ and at a given $X$, respectively.

As our first attempt to constrain $X$, we used the calibrated Gaia parallax of each PRE burster to determine which value of $X$ is more likely for the burster -- 0.73 or 0.
We calculated the Bayes factor $K=K^{X=0.7}_{X=0}$ using

\begin{equation}
\label{eq:bayes_factor}
\begin{split}
K &= \frac{\mathrm{P}(\pi_1-\pi_0|X=0.7)}{\mathrm{P}(\pi_1-\pi_0|X=0)}\\
&= \frac{\int^{\infty}_{0} \frac{1}{\sigma_{0.7}}\exp\left[-\frac{1}{2}\left(\frac{1/D-\mu_{\pi}}{\sigma_{\pi}}\right)^2-\frac{1}{2}\left(\frac{D-D_{0.7}}{\sigma_{0.7}}\right)^2\right]\mathrm{d}D}
{\int^{\infty}_{0} \frac{1}{\sigma_{0}}\exp\left[-\frac{1}{2}\left(\frac{1/D-\mu_{\pi}}{\sigma_{\pi}}\right)^2-\frac{1}{2}\left(\frac{D-D_{0}}{\sigma_{0}}\right)^2\right] \mathrm{d}D},
\end{split}
\end{equation}
where $\pi_1-\pi_0=\mu_{\pi}\pm\sigma_{\pi}$, $D|_{X=0.7}=D_{0.7} \pm \sigma_{0.7}$ and $D|_{X=0}=D_{0} \pm \sigma_{0}$. 
For the calculation of $K^{X=0.7}_{X=0}$, we did not use any Galactic prior (see Equation~\ref{eq:posterior_distance_PDF} for explanation), as the extra constraint given by a Galactic prior is negligible when $X$ is fixed.
The results of $K^{X=0.7}_{X=0}$ are presented in Table~\ref{tab:constrain_composition}.

Among the three $\log_{10} K^{X=0.7}_{X=0}$ values, the one for \cen\ is the most offset from 0, which suggests the calibrated parallax of \cen\ substantially (when $0.5<|\log_{10}K^{X=0.7}_{X=0}|<1$, \citealp{Kass95}) favors $X=0$ over $X=0.7$. 
Merely 2 PRE bursts have ever been observed from \cen\ in 1969 \citep{Belian72} and 1979 \citep{Matsuoka80}. The long intervals between PRE bursts of \cen\ imply small accretion rates and low $X$ at the time of PRE bursts, despite hydrogen signatures observed from its donor star \citep[e.g.][]{Shahbaz14}. This expectation is confirmed by the $\log_{10} K^{X=0.7}_{X=0}$ of \cen.

It is noteworthy that hydrogen-poor nuclear fuel at the time of PRE bursts is also favored for \cyg\ and \uua. 
In fact, the previous independent test of the simplistic PRE model using GC PRE bursters also suggests hydrogen-poor nuclear fuel at the time of PRE bursts for the majority of the 12 GC PRE bursters \citep{Kuulkers03}.
The statistical inclination towards low $X$, if confirmed with a larger sample of PRE bursters, might be partly attributed to a selection effect explained as follows.

For each PRE burster, only the brightest PRE bursts are selected to calculate its nominal PRE distance \citep{Galloway08,Galloway20,Chevalier89}. This selection criterion results in a nominal PRE distance that tends to decrease with time as brighter bursts are discovered. 
This tendency is clearly shown by the fact that almost all the nominal PRE distances registered in Table~9 of \citet{Galloway08} are larger than their counterparts in Table~8 of \citet{Galloway20}. 
If we assume the variability of observed fluxes of PRE bursts is mainly caused by the fluctuation of $X$ (at the time of different PRE bursts), the selection criterion of PRE bursters would lead to an increasingly biased PRE-burst sample that is more likely low-$X$.
This selection effect would have a big impact on the $X$ of the frequent PRE burster \cyg\ (see $n_\mathrm{burst}$ in Table~1 of \citealp{Galloway20}), but less so for \uua\ and \cen. 
A low $X$ of \uua\ (if confirmed in the future) can be explained by the suggestion that the donor star of \uua\ is likely a helium white dwarf \citep{int-Zand05}.
On the other hand, the revised nominal PRE distances of PRE bursters \citep{Galloway20} relieve the tension between the pre-2008 nominal PRE distances (see the references in Table~1 of A21) and the Gaia DR2 distances (A21), as mentioned in Section~\ref{sec:intro}.

\begin{table}
\caption{Bayes factor $K=K^{X=0.7}_{X=0}=P(\pi_1-\pi_0|X=0.7)/P(\pi_1-\pi_0|X=0)$ (ratio of two conditional probabilities), where $X$ refers to the mass fraction of hydrogen of the nuclear fuel.}
\label{tab:constrain_composition}
\centering
\begin{tabular}{@{}ccccc@{}}
\hline\hline
PRE & $D(X\!=\!0)$ & $\log_{10} K$ & $D^\mathrm{post}$ & $X^\mathrm{post}$ \\
burster & (kpc) & & (kpc) & \\
\hline\
%\vspace{1cm}
\cen\ & 1.2(3)\,$^\mathrm{a}$ & $-0.83$ & $1.4(2)$ & $0.2^{+0.3}_{-0.1}$ \\
\cyg\ & 11.6(9)\,$^\mathrm{b}$ &  $-0.51$ & $11.3^{+0.9}_{-0.8}$ & $0.2^{+0.2}_{-0.1}$\\
\uua\ & 3.9(2)\,$^\mathrm{b}$ & $-0.39$ & $3.5(3)$ & $0.3^{+0.3}_{-0.2}$ \\
\hline\hline
\end{tabular}
\tabnote{$^\mathrm{a}$ \citet{Chevalier89}; $^\mathrm{b}$ \citet{Galloway20}.}
\end{table}

\begin{figure*}
    \centering
	\begin{tabular}{cc}
    \includegraphics[width=85mm]{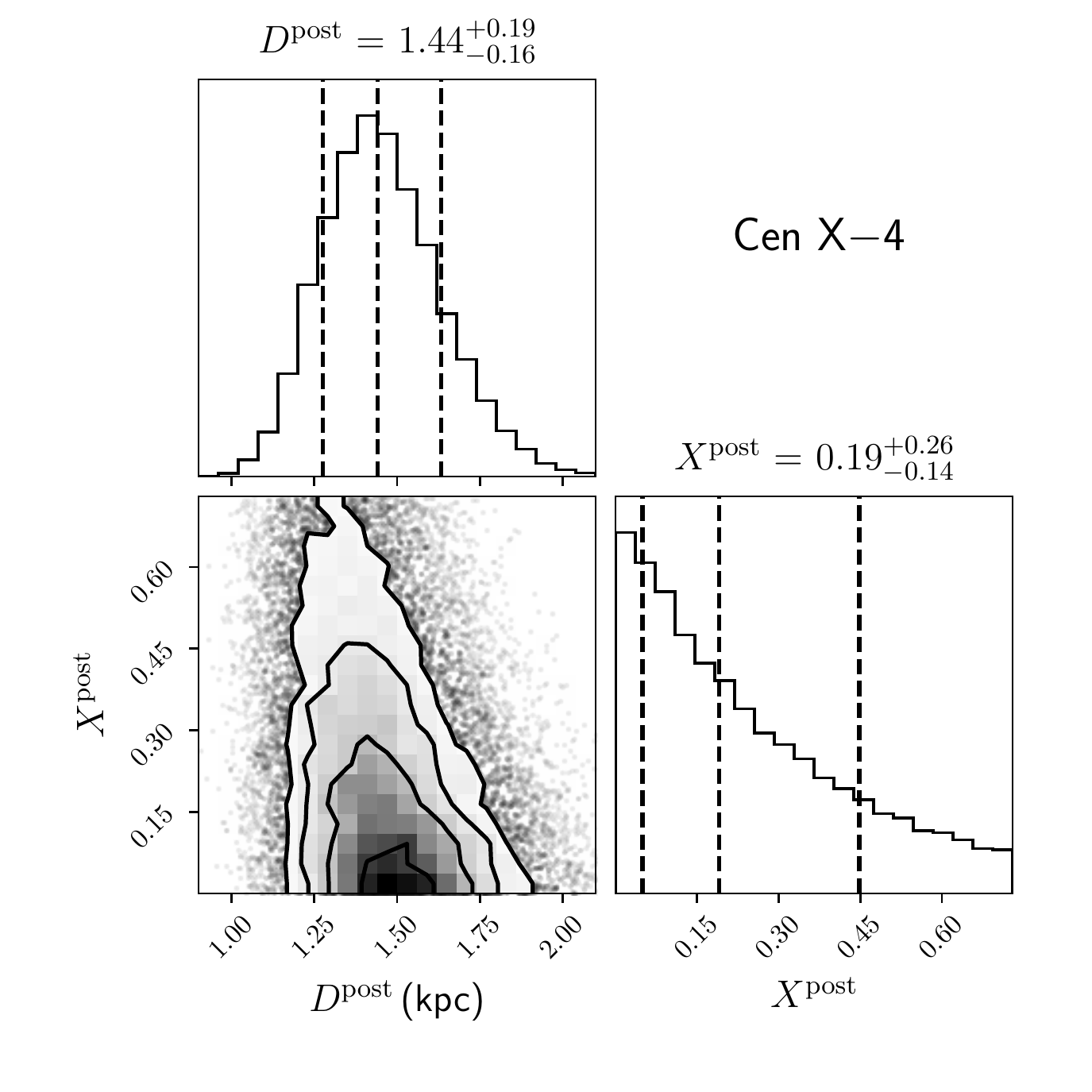} &   \includegraphics[width=85mm]{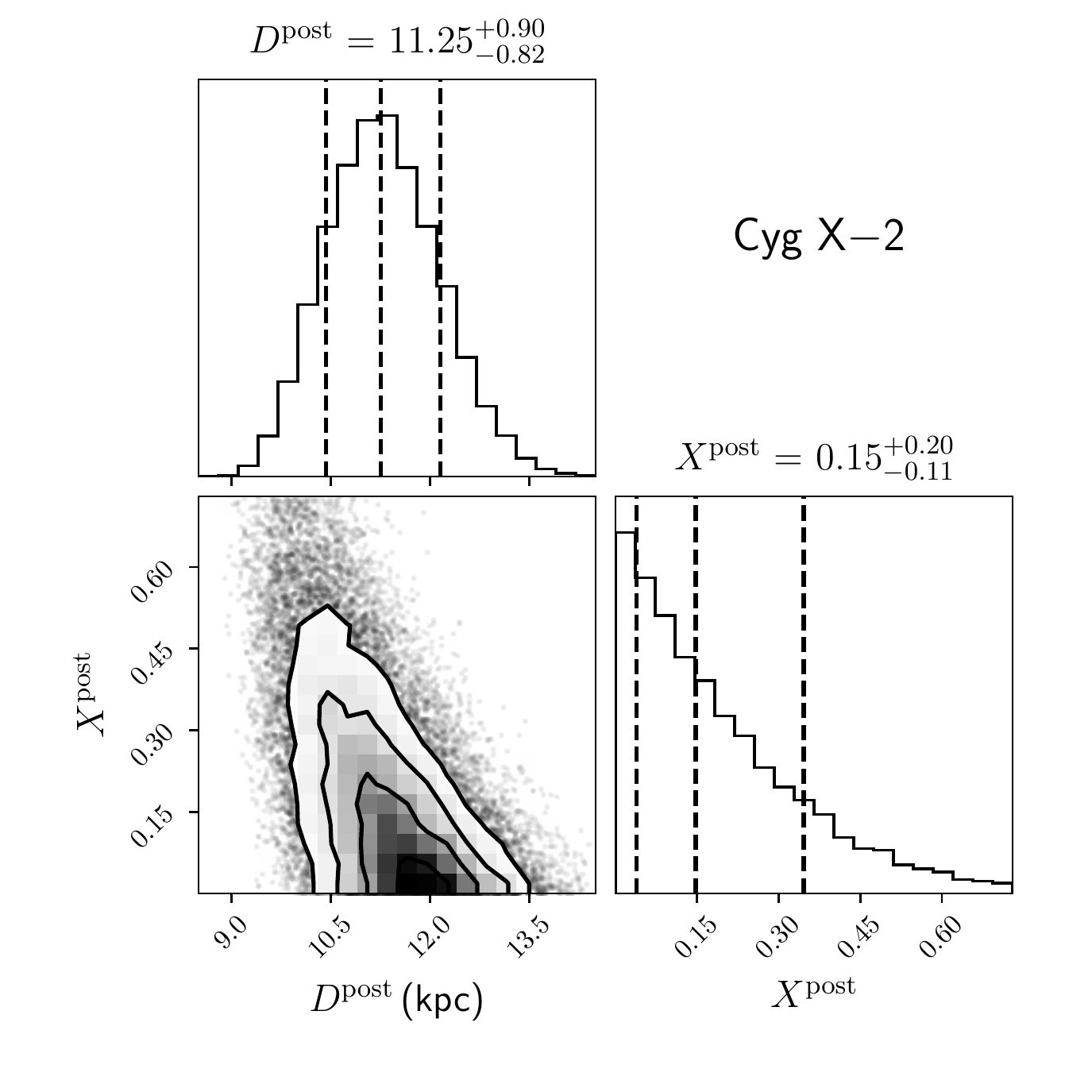} \\
    \multicolumn{2}{c}{\includegraphics[width=85mm]{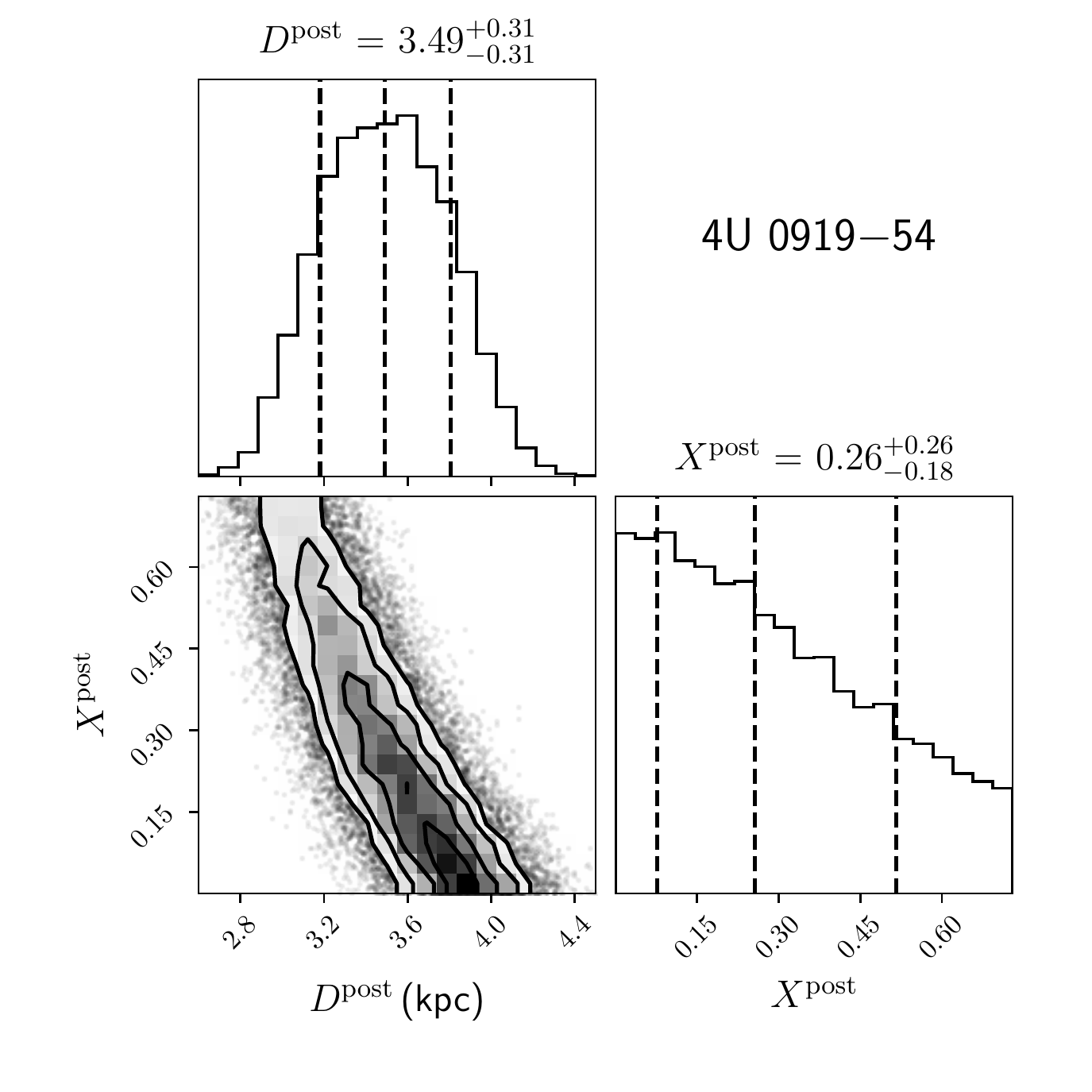} }\\
    \end{tabular}	

	\caption{2-D histograms and marginalized 1-D histograms for posterior distance $D^{\mathrm{post}}$ and posterior mass fraction of hydrogen of nuclear fuel $X^{\mathrm{post}}$  simulated with {\tt bilby} \citep{Ashton19} and plotted with {\tt corner.py} \citep{Foreman-Mackey16}. The $n$-th contour in each 2-D histogram contains $1-\exp\left(-n^2/2\right)$ of the simulated sample \citep{Foreman-Mackey16}.
	The vertical lines in the middle and two sides mark the median and central 68\% of the sample, respectively.}
    \label{fig:posterior_distances}
\end{figure*}

\subsubsection{Refining nominal PRE distances and compositions of nuclear fuel}
\label{subsubsec:post_D_and_X_assuming_PRE_model_is_right}
To better understand the statistical properties of $X$ (and to refine the PRE distance), we subsequently constrained both $X$ and the nominal PRE distance $D$ of each PRE burster with its calibrated parallax in a Bayesian way. The 2-D joint probability distribution (JPD) we applied is
\begin{equation}
\label{eq:posterior_distance_PDF}
\begin{split}
    \phi &= \phi(D,X\ |\ \pi_1-\pi_0,D_0) \\
    &\propto \phi(\pi_1-\pi_0,D_0\ |\ D,X) \\
    &\propto
    \rho(D) \exp\left[{-\frac{1}{2}\left( \frac{\frac{1}{D}-\mu_{\pi}}{\sigma_\pi}\right)^2-\frac{1}{2}\left( \frac{D-\frac{D_0}{\sqrt{X+1}}}{\frac{\sigma_0}{\sqrt{X+1}}}\right)^2}\right], \\
    & (D>0\ \text{and}\ X \in [0,0.73]),
\end{split}
\end{equation}
where $\rho(D)$ represents the
Galactic prior term, i.e. the prior probability of finding a PRE burster at a distance $D$ along a given line of sight. We approached $\rho$ by the Galactic mass distribution of LMXBs, which can be calculated with Equation~5 of \citet{Atri19} using parameters provided by \citet{Grimm02,Atri19}. 
Among the three components of the Galactic mass density (i.e. $\rho_\mathrm{bulge}$, $\rho_\mathrm{disk}$ and $\rho_\mathrm{sphere}$, see \citealp{Grimm02} for explanations), the Galactic-disk component $\rho_\mathrm{disk}$ dominates the Galactic mass density budget for LMXBs at the light of sight to any of the three PRE bursters. Therefore, for this work alone, we adopted $\rho_\mathrm{disk}$ (see Equation~5 of \citealp{Grimm02}) as the Galactic prior term $\rho$.

Following the JPD, we simulated posterior \{$D$, $X$\} pairs (hereafter referred to as $D^\mathrm{post}$ and $X^\mathrm{post}$), using {\tt bilby}\footnote{\url{https://lscsoft.docs.ligo.org/bilby/}} \citep{Ashton19}. 
For this simulation, we adopted a uniform distribution between 0 and 0.73 for $X$ and a uniform distribution between 0 and 50\,kpc for $D$ (where using a larger upper limit of $D$ does not change the result).
Based on the marginalized probability density functions (PDFs) of the simulated \{$D^\mathrm{post}$, $X^\mathrm{post}$\} chain, we estimated $D^\mathrm{post}$ and $X^\mathrm{post}$, which are presented in Figure~\ref{fig:posterior_distances} and Table~\ref{tab:constrain_composition}.

In general, the conversion of parallax to distance depends on the adopted Galactic prior \citep{Bailer-Jones15}. As an example, A21 found that the application of different Galactic priors \citep[e.g.][]{Bailer-Jones21,Atri19} results in inconsistent distances (see Figure~3 of A21). 
Unlike A21, we applied an extra constraint on $D$ provided by the PRE distance (see Equation~\ref{eq:posterior_distance_PDF}), which gives a handle on $X$ and improves the constraint on $D$.
With this extra constraint, the dependence on the Galactic prior becomes negligible for all of the three PRE bursters: their $D^\mathrm{post}$ and $X^\mathrm{post}$ stay almost the same without the $\rho(D)$ term.

The PDFs of $X$ provide richer information on $X$ than $K^{X=0.7}_{X=0}$. As expected from $K^{X=0.7}_{X=0}$, the PDF of $X$ for \cen\ favors a hydrogen-poor nature of nuclear fuel.
In addition to the new constraint we put on $X$, the nominal PRE distance of each PRE burster incorporating the parallax information is supposed to be more reliable (though not necessarily more precise). We note that the refined nominal PRE distances are still model-dependent; their accuracies are subject to the validity of the simplistic PRE model.

\subsection{A roadmap for testing the simplistic PRE model with parallax measurements}
\label{subsec:roadmap_for_model_testing}
The discussion in Section~\ref{subsec:posterior_distances} is based on the correctness of the simplistic PRE model. 
Now we discuss the potential feasibility of using parallaxes (of PRE bursters) to test the simplistic PRE model. 
We first define the PRE distance correction factor $\eta$ ($\eta>0$) using the relation 
\begin{equation}
\label{eq:eta}
D_X = \eta\ \frac{D_0}{\sqrt{X+1}},
\end{equation}
where $D_0$ stands for the nominal PRE distance at $X=0$ (derived from the simplistic PRE model), and $D_X$ refers to the ``true'' distance. If the simplistic PRE model is correct, then $\eta=1$. However, $\eta\ \neq\ 1$ is possible due to various causes including the stochasticity of PRE burst luminosity and anisotropy of PRE burst emission \citep[e.g.][]{Sztajno87}. The latter cause implies $\eta$ to be geometry-dependent (e.g. viewing-angle-related); if all PRE bursters have different geometry-dependent $\eta$, a search for a global deviation of $\eta$ from unity with a sample of PRE bursters would be less practical. Nevertheless, it remains practical to constrain $\eta$ on an individual basis. One way to make this constraint is using the generalized $X$, defined as
\begin{equation}
\label{eq:X_prime}
X'+1=\frac{X+1}{\eta^2}.
\end{equation}
Accordingly, $D_X=D_0/\sqrt{X'+1}$. Thanks to the same mathematical formalism, we can re-use Equation~\ref{eq:posterior_distance_PDF} as the JPD for statistical analysis, except that the domain of $X'$ is widened to $X'>-1$.

\begin{figure*}
    \centering
    \begin{tabular}{cc}
    \includegraphics[width=85mm]{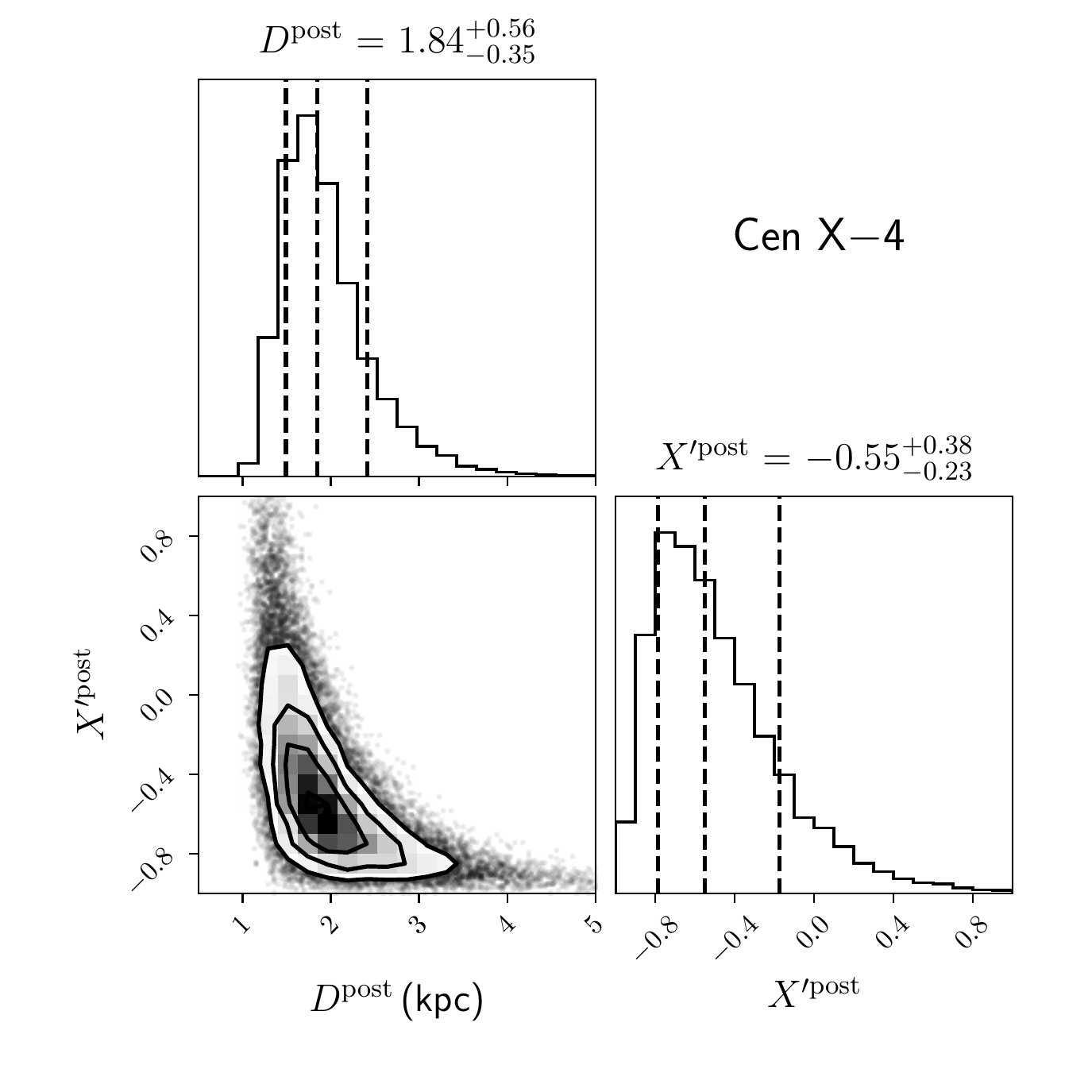} &   \includegraphics[width=85mm]{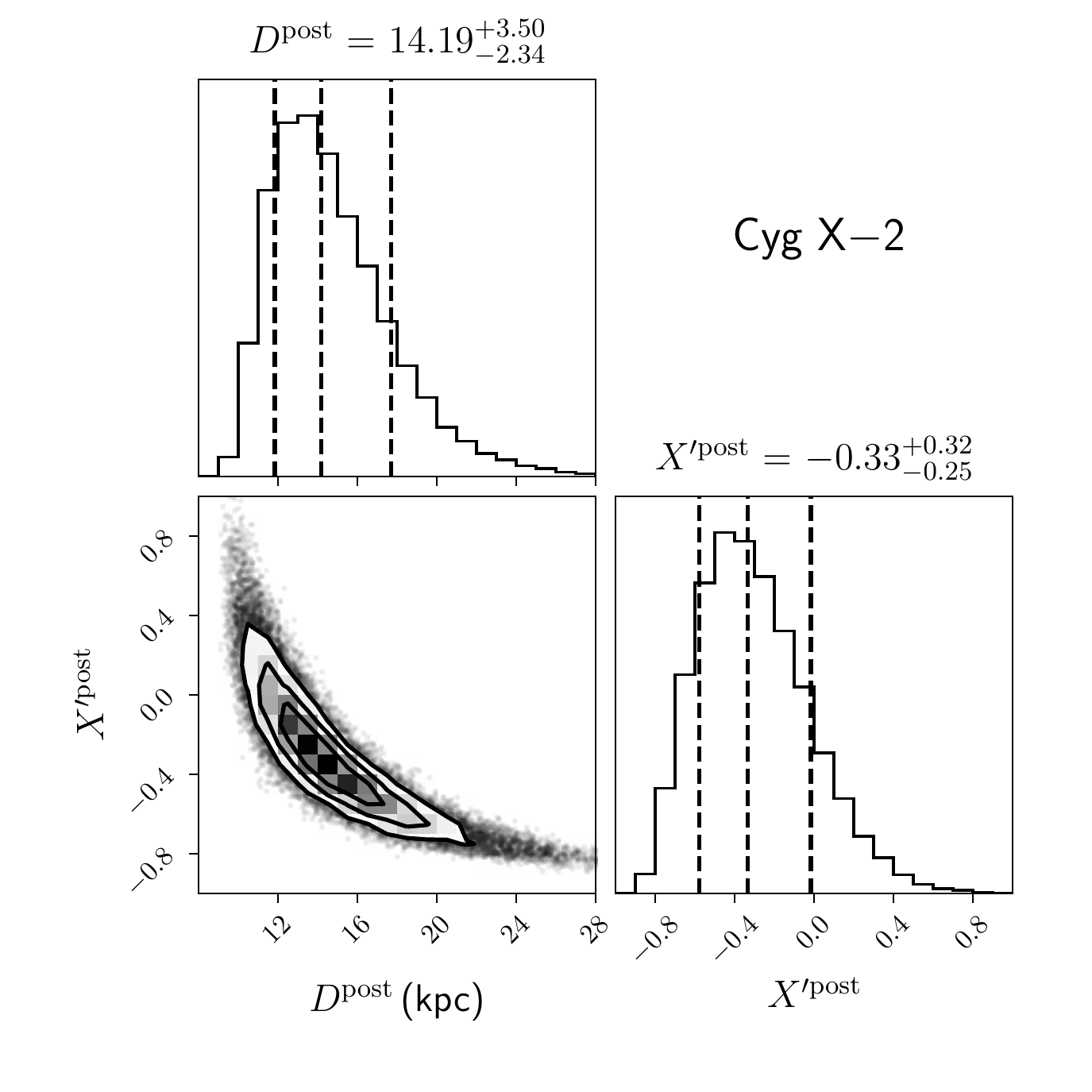} \\
    \multicolumn{2}{c}{\includegraphics[width=85mm]{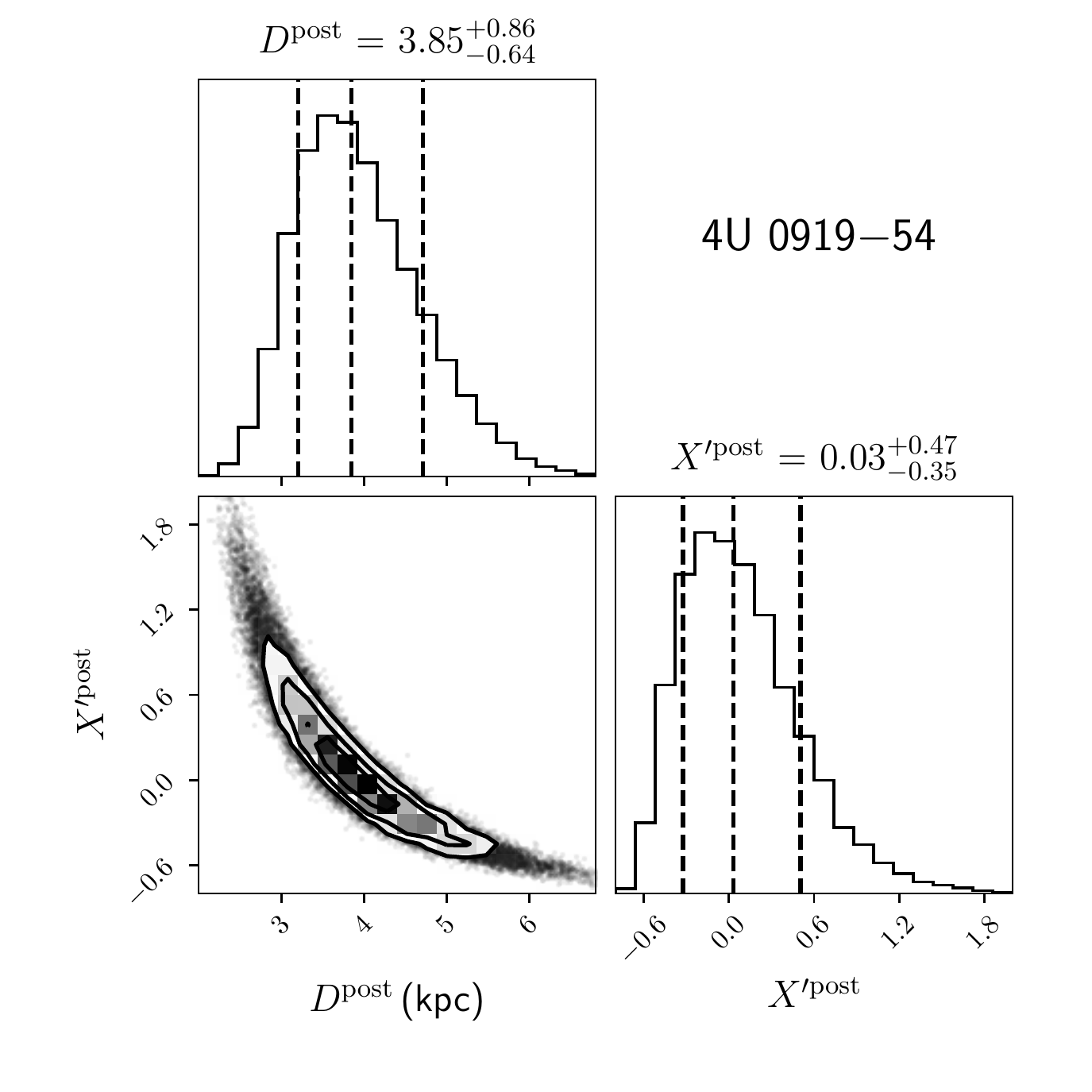} }\\
    \end{tabular}	

	\caption{2-D histograms and marginalized 1-D histograms of posterior distance $D^{\mathrm{post}}$ and $X'^{\mathrm{post}}$ (defined in Section~\ref{subsec:roadmap_for_model_testing}) simulated with {\tt bilby} \citep{Ashton19} and plotted with {\tt corner.py} \citep{Foreman-Mackey16}. The $n$-th contour in each 2-D histogram contains $1-\exp\left(-n^2/2\right)$ of the simulated sample \citep{Foreman-Mackey16}.
	The vertical lines in the middle and two sides mark the median and central 68\% of the sample, respectively.}
    \label{fig:posterior_X1_D}
\end{figure*}

The simulated \{$D^{\mathrm{post}}$, $X'^{\mathrm{post}}$\} chain (posterior $D$ and $X'$) is depicted in Figure~\ref{fig:posterior_X1_D}. 
If the simplistic PRE model is correct, then $\eta=1$; as a result, $X'$ is supposed to fall into the ``standard'' domain of [0, 0.73]. None of the three PDFs of $X'$ decisively rules out $X'$ from this standard domain. However, we notice $X'<0$ at 90\% confidence for \cen, which moderately disfavors the simplistic PRE model.

As the domain of $X'$ is broadened from that of $X$, the fractional precision of $D^{\mathrm{post}}$ becomes predominantly reliant on that of the parallax.
In this regime, $D^{\mathrm{post}}$ and $X'^{\mathrm{post}}$ become more dependent on the adopted Galactic prior.
To understand the degree of this dependence, we re-estimated $D^{\mathrm{post}}$ and $X'^{\mathrm{post}}$ after resetting the Galactic prior to $\rho(D)=1$, which are summarized in Table~\ref{tab:different_Galactic_prior}.
For \cyg\ with insignificant $\pi_1-\pi_0$, the dependence on the Galactic prior is evident, despite the consistency between the two sets of $D^{\mathrm{post}}$ and $X'^{\mathrm{post}}$.
For \cen\ and \uua\ both having relatively significant $\pi_1-\pi_0$, the degree of the dependence varies with the Galactic coordinates of the targets. In particular, we found the $D^{\mathrm{post}}$ and $X'^{\mathrm{post}}$ of \cen\ are robust against the selection of Galactic priors in Equation~\ref{eq:posterior_distance_PDF}. 

\begin{table}
\caption{$D^{\mathrm{post}}$ and $X'^{\mathrm{post}}$ (see Section~\ref{subsec:roadmap_for_model_testing}) inferred from the the Galactic prior $\rho(D)$ of Equation~\ref{eq:posterior_distance_PDF}, in comparison to the counterparts (noted as $\tilde{D}^{\mathrm{post}}$ and $\tilde{X'}^{\mathrm{post}}$) given $\rho(D)=1$.}
\label{tab:different_Galactic_prior}
\centering
\begin{tabular}{@{}ccccc@{}}
\hline\hline
PRE & $D^{\mathrm{post}}$ & $\tilde{D}^{\mathrm{post}}$ & $X'^{\mathrm{post}}$ & $\tilde{X'}^{\mathrm{post}}$ \\
burster & (kpc) & (kpc) &  &  \\
\hline\
%\vspace{1cm}
\cen\ & $1.8^{+0.6}_{-0.4}$ & $1.8^{+0.6}_{-0.3}$ & $-0.55^{+0.38}_{-0.23}$ & $-0.54^{+0.38}_{-0.24}$  \\
\cyg\ & $14^{+4}_{-2}$ & $18^{+13}_{-6}$ & $-0.3(3)$ & $-0.6^{+0.5}_{-0.3}$ \\
\uua\ & $3.9^{+0.9}_{-0.6}$ & $4.1^{+1.2}_{-0.8}$ & $0.0^{+0.5}_{-0.4}$ & $-0.1^{+0.5}_{-0.4}$\\
\hline\hline
\end{tabular}
\end{table}

Using $X'$ as a probe of the simplistic PRE model has the advantage of being independent of the PDF for $X$, which is usually not available. 
Future $D_0$ and $\pi_1-\pi_0$ measurements with higher precision will sharpen the above test of the simplistic PRE model. Lower fractional uncertainty of $\pi_1-\pi_0$ will not only reduce the strong correlation between $D^{\mathrm{post}}$ and $X'^{\mathrm{post}}$ for \cyg\ and \uua\ (shown in Figure~\ref{fig:posterior_X1_D}), but also lessen the dependence of inferred $D^{\mathrm{post}}$ and $X'^{\mathrm{post}}$ on the adopted Galactic prior.
On the other hand, constraining $X$ at PRE bursts independently (with, for example, ignition models of X-ray bursts) is the key to singling out $\eta$. Given that the simplistic PRE model is not significantly disfavored by any of the 3 PRE bursters, the refined distances in Table~\ref{tab:constrain_composition} (based on the correctness of the simplistic PRE model) are still valid.

\subsection{Future prospects}
\label{subsec:future_prospects}
The potential of constraining $X$ and testing the simplistic PRE model using Gaia parallaxes of PRE bursters hinges on the achievable parallax uncertainties. 
Future Gaia data releases include Gaia Data Release 4 (DR4) (for the five-year data collection) and the data release after extended observations of up to 5.4 years (\url{https://www.cosmos.esa.int/web/gaia/release}).
We predict the achievable parallax ($\pi_1$) and proper motion uncertainties of \cen\ for future Gaia data releases using the simulation method detailed in Appendix~\ref{app:simulating_parallax_precision}. 
In the same way, we estimated the predicted $\pi_1$ uncertainty $\hat{\sigma}_{\pi_1}$ of \cyg, \uua\ and \xb\ for future data releases and DR2, which are listed in Table~\ref{tab:simulated_parallax_precision}. 
The predicted $\pi_1$ uncertainties for DR2, denoted as $\hat{\sigma}^\mathrm{DR2}_{\pi_1}$, generally agree with the real values $\sigma^\mathrm{DR2}_{\pi_1}$ (see Table~\ref{tab:simulated_parallax_precision}).
In general, our simulation shows Gaia parallax and proper motion precision improves with observation time $t$ at roughly $t^{-1/2}$ and $t^{-3/2}$, respectively.

\begin{table}
\caption{Predicted parallax ($\pi_1$) uncertainties $\hat{\sigma}_{\pi_1}$ for \cen, \cyg\ and \uua\ obtained with simulations (see Appendix~\ref{app:simulating_parallax_precision} for details). For comparison, real values of $\sigma^\mathrm{DR2}_{\pi_1}$ are provided as $\sigma^\mathrm{DR2}_{\pi_1}$.}
\label{tab:simulated_parallax_precision}
\centering
\begin{tabular}{@{}cccccc@{}}
\hline\hline
PRE & $\hat{\sigma}^\mathrm{DR2}_{\pi_1}$ & $\sigma^\mathrm{DR2}_{\pi_1}$ & $\sigma^\mathrm{EDR3}_{\pi_1}$ & $\hat{\sigma}^\mathrm{DR4}_{\pi_1}$ & $\hat{\sigma}^\mathrm{ext}_{\pi_1}$ $^*$\\
burster & (mas) & (mas) & (mas) & (mas ) & (mas)\\
\hline\
%\vspace{1cm}
\cen\ & 0.16 & 0.15 & 0.13 & 0.10 & 0.07 \\
\cyg\ & 0.024 & 0.024 & 0.019 & 0.014 & 0.010\\
\uua\ & 0.08 & 0.07 & 0.06 & 0.05 & 0.03\\
\xb\ & 0.10 & 0.09 & 0.08 & 0.06 & 0.04\\
\hline\hline
\end{tabular}
\tabnote{$^*$ we adopt an indicative 5-yr extension on top of the Gaia 5-yr mission.}
\end{table}

According to Table~\ref{tab:zero_parallax_points}, the future parallax uncertainty of \cyg\ will be predominantly limited by the uncertainty of $\pi_0$, or $\sigma_{\pi_0}$. For this target, improvement in the estimation of the zero-point parallax correction and its uncertainty -- which is challenging due to the relative brightness of the source -- is essential.
In comparison, the $\sigma_{\pi_1}$ of \cen, \uua\ and \xb\ dominates the respective error budget of the calibrated parallax $\pi_1-\pi_0$. Therefore, one can expect $\pi_1-\pi_0$ of \cen, \uua\ and \xb\ to be potentially 1.9 times more precise (than EDR3) at the end of the extended Gaia mission (with 10-yr data), which would promise better constraints on $X$ and $X'$ for \cen\ and \uua.
On the other hand, to constrain $X$ and $X'$ for \xb\ with its already relatively precise Gaia parallax will require more PRE bursts to be detected from \xb.
Apart from Gaia astrometry, radio observations of PRE bursters during their outbursts using very long baseline interferometry technique can also potentially measure geometric parallaxes of PRE bursters, hence serving the same scientific goals outlined in this paper.

This work only deals with the simplistic PRE model, the hitherto most impactful PRE model. As is mentioned in Section~\ref{sec:intro}, PRE models are not the only pathway to the distances of type I X-ray bursters. Bayesian inference based on a burst ignition model \citep{Cumming00} was recently realized to estimate parameters including $X$ and $D$ \citep{Goodwin19}. 
We believe such inferences will be made for more type I X-ray bursters in the near future.
In Bayesian analysis based on burst ignition models \citep[e.g.][]{Cumming00,Woosley04}, model-independent geometric parallaxes will serve as prior knowledge and help refine the inferred parameters including $D$ and $X$; the parallaxes can also judge which burst ignition model is more likely correct with a Bayes factor analysis.

Finally, we reiterate that we have assumed $M_\mathrm{NS}=1.4\,\msun$ and $R_\mathrm{NS}=11.2$\,km for all PRE bursters. Therefore, strictly speaking, $\eta$ should also reflect the deviation of $M_\mathrm{NS}$ and $R_\mathrm{NS}$ from their respective assumed values. 
Though a large deviation of $M_\mathrm{NS}$ from 1.4\,\msun\ is rare, it is not impossible \citep{Shao20}. 
In general, $L_{\mathrm{Edd},\infty}$ would increase with larger $M_\mathrm{NS}$ and $R_\mathrm{NS}$ (see Equation~\ref{eq:Eddington_luminosity}).
For example, $L_{\mathrm{Edd},\infty}(M_\mathrm{NS}=2, R_\mathrm{NS}=15)$ is 40\% greater than $L_{\mathrm{Edd},\infty}(M_\mathrm{NS}=1.4, R_\mathrm{NS}=11.2)$. 
Such a $L_{\mathrm{Edd},\infty}$ would correspond to $X'=-0.29$ assuming $X=0$ and the simplistic PRE model is correct. This value is consistent with $X'=-0.55^{+0.38}_{-0.23}$ for \cen. 
Given that the NS mass distribution is relatively well constrained compared to the NS radius distribution, it is possible to constrain the NS radius of a PRE burster in a Bayesian framework (slightly different from this work) provided prior knowledge of $X$, $D$ and $M_\mathrm{NS}$, assuming the simplistic PRE model is correct.

\section{Conclusion}
\label{sec:conclusion}
This work joins the few previous efforts to determine the Gaia parallax zero-point $\pi_0$ for a single source using nearby (on the sky) background quasars.
We provide a new template for $\pi_0$ determination (which also includes the $\pi_0$ uncertainty) approached by the weighted standard deviation of quasar parallaxes with respect to the filter parameters (see Section~\ref{sec:parallax_calibration}). 
This work focuses on the small sample of PRE bursters with detected Gaia parallaxes $\pi_1$. A future work will apply the new template to a larger sample of targets, which can enable a comprehensive comparison with zero-parallax points estimated in other ways.

We also introduce a Bayesian framework to test the simplistic PRE model with parallaxes of PRE bursters on an individual basis. 
Though no parallax of the PRE bursters decisively disfavours the simplistic PRE model, the model is disfavored by the current \cen\ parallax at 90\% confidence.
At the end of the extended Gaia mission with 10-yr data, we expect the parallax precision to improve by a factor of 1.9, which will sharpen the test of the simplistic PRE model.
This test will also considerably benefit from an independent determination of $X$ (hydrogen mass fraction of nuclear fuel), which we highly encourage.

\section*{acknowledgements}
The authors thank Ilya Mandel, Chris Flynn and Adelle Goodwin for useful discussions, and are grateful to the anonymous referee for the helpful comments on the manuscript.
H.D. is supported by the ACAMAR (Australia-ChinA ConsortiuM for Astrophysical Research) scholarship, which is partly funded by the China Scholarship Council.
A.T.D is the recipient of an ARC Future Fellowship (FT150100415).
Parts of this research were conducted by the Australian Research Council Centre of Excellence for Gravitational Wave Discovery (OzGrav), through project number CE170100004.
This work has made use of data from the European Space Agency
(ESA) mission Gaia (\url{https://www.cosmos.esa.int/gaia}), processed by
the Gaia Data Processing and Analysis Consortium (DPAC, \url{https://www.cosmos.esa.int/web/gaia/dpac/consortium}).
Data analysis was partly performed on OzSTAR, the Swinburne-based supercomputer.
%\end{acknowledgements}

\begin{appendix}
\section{Simulating the parallax precision of \cen\ in future Gaia data releases}
\label{app:simulating_parallax_precision}
To simulate the uncalibrated parallax ($\pi_1$) precision of \cen\ for a future Gaia data release, we made two assumptions: {\bf 1)} \cen\ is observed by Gaia at equal intervals; 
{\bf 2)} $\sigma^i_k=\sigma^j_k$, ($k=\alpha, \delta$ and $i \neq j$), where $\sigma^i_\alpha$ and $\sigma^i_\delta$ stand for, respectively, the uncertainty of right ascension (RA) and declination at the $i$-th observation. 

The observation interval is determined with Table~1 of \citet{Gaia-Collaboration16}. At the ecliptic latitude of \cen\ ($-14\fdg1$), the number of observations is 63 over 5\,yr, corresponding to an interval of 29\,d.
It appears from Table~\ref{tab:before_calibration} that, for a specific target, $\sigma_\alpha/\sigma_\delta \approx \sigma_{\mu_\alpha}/\sigma_{\mu_\delta}$ (where $\sigma_{\mu_\alpha}$ and $\sigma_{\mu_\delta}$ represent uncertainty of $\mu_\alpha$ and $\mu_\delta$ respectively), which implies the ratio of spatial resolution between RA and declination deviates slightly from unity and changes with pointing. 
Hence, we set $\sigma^i_\alpha / \sigma^i_\delta$ to 1.21, which is the $\sigma_{\mu_\alpha}/\sigma_{\mu_\delta}$ of \cen\ from EDR3.

Simulated positions of \cen\ were generated with {\tt pmpar} (available at \url{https://github.com/walterfb/pmpar}) using the ``predictor mode'', based on the astrometric parameters of \cen\ in Table~\ref{tab:before_calibration}. 
We first simulated positions from MJD~56863 to MJD~57901 (the observing period of EDR3). The sole variable $\sigma^i_\alpha$ (and $\sigma^i_\delta$ accordingly) was tuned to make the parallax precision equal to 0.13\,mas, the current parallax precision of \cen. Subsequently, we stuck to the position uncertainties while extending the observing period to MJD~58689 (around the end day of the five-year period) and MJD~60516 (an indicative day 10 years after the first light of Gaia), which, respectively, yield parallax uncertainties of 0.095\,mas and 0.067\,mas.
Thus, we predict the parallax precision of \cen\ to improve by a factor of 1.3 and 1.9, respectively, with DR4 and 10-yr data.
Besides, we expect the proper motion precision of \cen\ to be enhanced by 2.7 and 6.7 times, respectively, with DR4 and 10-yr data.
\end{appendix}

\bibliographystyle{pasa-mnras}
\bibliography{haoding,GX17+2,PRE_bursters_Gaia_counterparts}

\end{document}